\documentclass[12pt]{article}
\usepackage{graphicx}
\usepackage{amssymb,mathtools}
\usepackage{amsmath}
\usepackage{color}
\usepackage{tabularx}
\usepackage{enumitem}
\usepackage{natbib}
\usepackage{hyperref}
\usepackage{dsfont}
\usepackage{algorithm}
\usepackage{algpseudocode}
\usepackage{multirow}
\usepackage{threeparttable}
\usepackage{lscape}
\usepackage{booktabs}
\usepackage{setspace}
\usepackage{url}
\usepackage{caption2}

\newcommand{\blind}{1}

\makeatletter
\newenvironment{breakablealgorithm}
  {
   \begin{center}
     \refstepcounter{algorithm}
     \hrule height.8pt depth0pt \kern2pt
     \renewcommand{\caption}[2][\relax]{
       {\raggedright\textbf{\ALG@name~\thealgorithm} ##2\par}%
       \ifx\relax##1\relax 
         \addcontentsline{loa}{algorithm}{\protect\numberline{\thealgorithm}##2}%
       \else 
         \addcontentsline{loa}{algorithm}{\protect\numberline{\thealgorithm}##1}%
       \fi
       \kern2pt\hrule\kern2pt
     }
  }{
     \kern2pt\hrule\relax
   \end{center}
  }
\makeatother

\newcommand{\abs}[1]{\left\lvert#1\right\rvert}
\newcommand{\norm}[1]{\left\lVert#1\right\rVert}

\global\long\def\prob{\mathbb{P}}

\newtheorem{thm}{Theorem}

\begin{document}

\def\spacingset#1{\renewcommand{\baselinestretch}%
{#1}\small\normalsize} \spacingset{1}


\if1\blind
{
  \title{\bf Computationally Efficient Whole-Genome Signal Region Detection for Quantitative and Binary Traits}
  \author{Fan Wang \\
    Department of Biostatistics, Mailman School of Public Health, \\Columbia University, New York, U.S.A\\
    Wei Zhang\\
    School of Mathematical Sciences, Center for Statistical Science, \\Peking University, Beijing, China\\
    Fang Yao\thanks{
     Fan Wang and Wei Zhang are co-first authors, and  Fang Yao is the corresponding author. E-mail: fyao@math.pku.edu.cn. This research is supported in part  by  the National Key R\&D Program of China (No. 2022YFA1003801), the National Natural Science Foundation of China (No. 12292981, 11931001), the LMAM and the Fundamental Research Funds for the Central Universities, Peking University. This research has been conducted using the UK Biobank Resource under project 79237.\hspace{.2cm}} \\
    School of Mathematical Sciences, Center for Statistical Science, \\Peking University, Beijing, China}
  \maketitle
} \fi

\if0\blind
{
  \bigskip
  \bigskip
  \bigskip
  \begin{center}
    {\LARGE\bf Computationally Efficient Whole-Genome Signal Region Detection for Quantitative and Binary Traits}
\end{center}
  \medskip
} \fi

\begin{abstract}
The identification of genetic signal regions in the human genome is critical for understanding the genetic architecture of complex traits and diseases. Numerous methods based on scan algorithms (i.e. QSCAN, SCANG, SCANG-STARR) have been developed to allow dynamic window sizes in whole-genome association studies. Beyond scan algorithms, we have recently developed the binary and re-search (BiRS) algorithm, which is more computationally efficient than scan-based methods and exhibits superior statistical power. However, the BiRS algorithm is based on two-sample mean test for binary traits, not accounting for multidimensional covariates or handling test statistics for non-binary outcomes. In this work, we present a distributed version of the BiRS algorithm (dBiRS) that incorporate a new infinity-norm test statistic based on summary statistics computed from a generalized linear model. The dBiRS algorithm accommodates regression-based statistics, allowing for the adjustment of covariates and the testing of both continuous and binary outcomes. 
This new framework enables parallel computing of block-wise results by aggregation through a central machine to ensure both detection accuracy and computational efficiency, and has 
theoretical guarantees for controlling family-wise error rates and false discovery rates while maintaining the power advantages of the original algorithm.
Applying dBiRS to detect genetic regions associated with fluid intelligence and prospective memory using whole-exome sequencing data from the UK Biobank, we validate previous findings and identify numerous novel rare variants near newly implicated genes. These discoveries offer valuable insights into the genetic basis of cognitive performance and neurodegenerative disorders, highlighting the potential of dBiRS as a scalable and powerful tool for whole-genome signal region detection.
\end{abstract}

\noindent%
{\it Keywords:} family-wise error rate, signal detection, distributed learning, whole genome association studies
\vfill

\newpage 
\spacingset{1.9} 
\section{Introduction}
\label{sec:intro}

Understanding the genetic underpinnings of human diseases and traits remains a central focus of genetic research. Genome-Wide Association Studies (GWAS) have been instrumental in exploring the genetic architecture of complex diseases and traits over the past decade \citep{Visscher17}. By using array-based technologies, GWAS analyzes millions of single nucleotide polymorphisms (SNPs) across the genome to identify those associated with specific traits or disease outcomes. While GWAS has successfully identified thousands of common genetic variants linked to disease susceptibility, these common variants explain only a small fraction of heritability, which is often referred to as the ``missing heritability problem''\citep{manolio2009finding}. The majority of genetic variants in the human genome are rare, and these rare variants are believed to contribute significantly to the unexplained heritability. However, the classical GWAS approach which focuses on single-SNP-based analysis has very limited power for analyzing rare variants, as the effect of a single SNP can be too small to be detected. \citep{liu2010versatile, bakshi2016fast}.

To address this limitation, whole genome sequencing (WGS) studies are being conducted to identify rare variants associated with disease susceptibility. Progress has been made in developing set-based association methods, which test multiple variants jointly by aggregating their effects within defined genomic regions. These methods include burden tests \citep{MORGENTHALER200728, Madsen2009}, the Sequence Kernel Association Test (SKAT) \citep{Wu2011AJHG}, and STAAR which incorporates variant functional annotations to enhance detection power \citep{Li2022nature}. The STAAR-O test extends the STAAR framework as an omnibus test by combining multiple annotation-weighted methods into a single unified test. A common challenge in these approaches is defining regions for variant sets, especially in non-coding or intergenic regions without clear functional boundaries. STAAR addresses this issue by applying gene-centric analysis to well-defined gene-associated regions and fixed sliding window-based analysis to regions without clear boundaries. Alternative approaches include the scan statistic \citep{Naus1982}, which systematically searches the human genome using a fixed window size. Mean-based scan statistic methods are later proposed for DNA copy number analysis, allowing the use of multiple window sizes in settings closely related to change-point detection problems \citep{Cai10, Olshen04, Zhang10}. Other notable frameworks, such as those based on the knockoff (i.e.KnockoffScreen) \citep{he2021identification, he2021genome}, conducts genome-wide set-based analyses to identify signal regions while mitigating the effects of correlation confounding. However, fixed-window approaches can result in a loss of power because the sizes of signal regions can vary across the genome. 

Further advancements have been made to enable signal region detection with dynamic window sizes. \cite{Li2020dynamic} introduced SCANG which combines the scan algorithm proposed by \cite{Cai10} with burden tests, SKAT, and the omnibus test to continuously scan the genome. However, this approach lacks comprehensive theoretical and empirical analysis of false discoveries and shows limited power when functional annotations are not incorporated \citep{Li2022nature}. Building on the scan algorithm, \cite{Lin20} developed the quadratic scan statistic (QSCAN), which aggregates information across intervals of varying sizes. QSCAN provides theoretical guarantees for controlling false discoveries and has demonstrated strong detection performance, particularly when signal regions contain both causal and neutral variants. However, scan-based methods require calculating test statistics for many candidate intervals and applying a fixed threshold for selection, which is computationally intensive and tends to be conservative. SCANG-STAAR \citep{Li2022nature} extends SCANG's dynamic window scanning by incorporating multiple functional annotations via STAAR to boost the power of detecting rare variant associations. While functional annotations can enhance power and interpretability, their effectiveness relies heavily on the quality and relevance of the annotations, which may introduce biases, computational challenges, and increased resource demand. 

Beyond the scan algorithm, \cite{zhang2023binary} proposed the binary and re-search (BiRS) algorithm for detecting signal regions with dynamic window sizes in whole genome sequencing (WGS) studies. The BiRS algorithm iteratively splits identified regions until a minimum size is reached, providing theoretical guarantees for FWER and FDR, which is shown more computationally efficient than scan-based algorithms and demonstrates superior power compared to QSCAN and KnockoffScreen. Impressively, BiRS surpasses SCANG-STAAR in detecting moderate or weak signals without requiring functional annotations. The combination of computational speed, increased detection power and theoretical rigor make BiRS as a significant advancement in signal region detection for WGS studies. The BiRS algorithm was combined with the DCF two-sample mean test \citep{Xue20} for direct application to binary traits. However, it has not yet been extended to correct for multi-dimensional covariates or to test for non-binary outcomes.

In this paper, we generalize the BiRS algorithm to a distributed version (dBiRS) to combine with a new infinity-norm test statistic based on the summary statistics obtained from a generalized linear model. As the proposed test statistic fundamentally differs from the DCF two-sample mean test, this extension is non-trivial. We develope new theoretical guarantees to ensure the size control of dBiRS while preserving the detection accuracy of the original BiRS. The dBiRS algorithm operates in two main stages, combining results from local and central machines to support parallel computing and maintain computational efficiency, similar to distributed learning frameworks  \citep{cai2022distributed}. In the first stage, BiRS is applied within genomic blocks, where local machines detect signal regions by calculating test statistics and thresholds within each block. Instead of simply aggregating block-wise results using a global threshold, the detected signal regions, along with their corresponding test statistics and thresholds, are transferred to a central machine. In the second stage, the central machine evaluates the significance of each block by performing a second layer of BiRS based on block-wise infinity-norm test statistics. Finally, a new threshold is constructed by multiplier bootstrap to reassess the signal regions within the significant blocks. We have shown theoretically that the dBiRS algorithm is able to consistently identifies true signal regions under more general alternative structures and in the presence of model misspecification under the alternative hypothesis. Simulated studies also demonstrate that dBiRS is more accurate and robust than the state-of-the-art KnockoffScreen and scan procedures.

Finally, we apply the BiRS algorithm to analyze whole exome sequence (WES) data from the UK Biobank, aiming to identify signal regions for intelligence and prospective memory. The dBiRS algorthm identified signal regions involving 327 genes, including 84 of which were previously reported to be assocaited with intelligence. Notable discoveries include both common and rare variants linked to cognitive functions, such as those in COL16A1, CRTC2, and PTPRF. Variants in genes like CRTC2, BRWD1, and TOP2B were also identified, highlighting their roles in endothelial function, immune system regulation, and neuronal development, respectively. In addition, the analysis revealed rare variants in 22 genes not previously associated with intelligence. Some of these genes are linked to Alzheimer’s disease, brain connectivity, and neuronal development, providing novel insights into the genetic basis of cognitive traits. For prospective memory, the study identified eight novel genes, including OMA1 and CNGB3, which are associated with neuroimaging measurements and cognitive functions.

The rest of the article is organized as follows. In Section \ref{sec:method}, we introduce the testing procedure under GLM and describe the proposed dBiRS algorithm. We conduct comprehensive simulations in Section \ref{sec:simulation} to demonstrate that the proposed method enjoys preferably numerical performance compared with existing approaches. In Section \ref{sec:application}, we apply the dBiRS algorithm to conduct Whole Exome Sequencing (WES) analyses on intelligence and prospective memory, aiming to deepen our understanding of cognitive aging and uncover genetic factors contributing to the risk of neurodegenerative disorders. We conclude the article with a discussion in Section \ref{sec:discussion}, while the theoretical properties of the proposed algorithm, including size control and detection consistency, are deferred to the Appendix. Technical assumptions, lemmas, and proofs of the theoretical results, along with additional simulation and application results, are provided in the Supplementary Material.

\section{Distributed detection algorithm}
\label{sec:method}
\subsection{Global test}
\label{subsec:global}
Suppose there are $n$ observations in the study. For the $i$-th observation, $Y_i$ represents the outcome, $X_i = (X_{i1}, \dots, X_{iq})^\top$ is a vector containing $q$ covariates, and $G_i = (G_{i1}, \dots, G_{ip})^\top$ is the genotype vector with $p$ variants. Let $Y = (Y_1, \dots, Y_n)^\top$, $\boldsymbol{X} = (X_1, \dots, X_n)^\top$, and $\boldsymbol{G} = (G_1, \dots, G_n)^\top$. Conditional on $X_i$ and $G_{i}$, we assume that $Y_i$ belongs to an exponential family with the density $
f(Y_i) = \exp\left\{\frac{Y_i \theta_i - b(\theta_i)}{a_i(\phi)} + c(Y_i, \phi)\right\},
$
where $a_i(\cdot)$, $b(\cdot)$, and $c(\cdot)$ are known functions, and $\theta_i$ and $\phi$ are the canonical parameter and dispersion parameter, respectively, which indicates that $Y$ following a generalized linear model (GLM):
\begin{equation}
\label{eqn:wgs_model1}
    g(\eta) = \boldsymbol{X}\gamma + \boldsymbol{G}\beta,
\end{equation}
where $\eta = \mathbb{E}(Y \mid \boldsymbol{X}, \boldsymbol{G})$ and $g(\cdot)$ is a monotone link function. 

Under the global null model where no genetic effect is present across the genome (i.e. $\beta=0$), the GLM in \eqref{eqn:wgs_model1} simplifies to $g(\eta) = \boldsymbol{X}\gamma$. Let $\hat{\eta}_0 = g^{-1}(\boldsymbol{X}\hat{\gamma})$, where $\hat{\gamma}$ is the maximum likelihood estimator (MLE) of $\gamma$ under the global null model. The variance of $Y_i$ is $\operatorname{var}\left(Y_i\right)=a_i(\phi) v\left(\eta_i\right)$, where $v\left(\eta_i\right)=b^{\prime \prime}\left(\theta_i\right)$ is a variance function. We define $\boldsymbol{\Lambda}=\operatorname{diag}\left\{a_1(\phi) v\left(\eta_{01}\right), \ldots, a_n(\phi) v\left(\eta_{0 n}\right)\right\}$ and let $\boldsymbol{P}=\boldsymbol{\Lambda}^{-1}-\boldsymbol{\Lambda}^{-1} \boldsymbol{X}\left(\boldsymbol{X}^T \boldsymbol{\Lambda}^{-1} \boldsymbol{X}\right)^{-1} \boldsymbol{X}^T \boldsymbol{\Lambda}^{-1}$. 

In genome-wide association studies (GWAS) and whole-genome sequencing (WGS) studies, the test statistic for the $j$-th variant is constructed using the working marginal model $
g\left(\eta_i\right) = X_i^T \alpha + G_{ij} \beta_j,$ where we regress $Y_i$ on each variant $G_{ij}$, adjusting for the covariates $X_i$.  The marginal score test statistic for $\beta_j$ of the $j$-th variant is given by
\begin{equation}
U_j = G_{.j}^T\left(Y - \hat{\eta_0}\right)/\sqrt{n}.
\end{equation}
Marginal score statistics $U_{j}$'s are often made available in public databases or provided by investigators to facilitate meta-analysis across multiple cohorts. Let $U=(U_{1},\dots,U_{p})^{\top}$, under the global null model (i.e. $\beta=0$), $U \sim \mathcal{N}(0, \boldsymbol{\Sigma})$, where $\boldsymbol{\Sigma}= \boldsymbol{G}^\top P \boldsymbol{G}/n$. Let $\boldsymbol{\hat{\Lambda}}=\operatorname{diag}\left\{a_1(\hat{\phi}) v\left(\hat{\eta}_{01}\right), \ldots, a_n(\hat{\phi}) v\left(\hat{\eta}_{0 n}\right)\right\}$, where $\hat{\phi}$ is the MLE of $\phi$ under the global null hypothesis and $\hat{\eta}_{0i}$ is the $i$-th coordinate of $\hat{\eta}_0$, $i = 1, \dots, n$. Let $\boldsymbol{\hat{P}}=\boldsymbol{\hat{\Lambda}}^{-1}-\boldsymbol{\hat{\Lambda}}^{-1} \boldsymbol{X}\left(\boldsymbol{X}^T \boldsymbol{\hat{\Lambda}}^{-1} \boldsymbol{X}\right)^{-1} \boldsymbol{X}^T \boldsymbol{\hat{\Lambda}}^{-1}$. In practice, $\Sigma$ can be well approximated by $\boldsymbol{G}^\top \hat{P} \boldsymbol{G}/n$.

To test for the global null hypothesis, our proposed test statistic is defined as the maxima of marginal scores:
\[
T = \norm{U}_{\infty} = \norm{\boldsymbol{G}^\top(Y - \hat{\eta}_0)/\sqrt{n}}_{\infty},
\]
where $\norm{x}_{\infty} = \max\left\{\abs{x_1}, \dots, \abs{x_p}\right\}$ for $x \in \mathbb{R}^p$. The null hypothesis is rejected at a specified significance level $\alpha$ if $T > c(\alpha)$, where $c(\alpha)$ is a predefined threshold, specifically, 
\begin{equation*}
    c(\alpha) = \inf\left\{ t \in \mathbb{R}: \mathbb{P}\left( T \leq t \right) \geq 1 - \alpha \right\}.
\end{equation*}

Computing the threshold $c(\alpha)$ requires the eigenvalues of $\hat{\Sigma}$, which are computationally intensive to calculate in practice when $p$ is large. To address this challenge, we propose an efficient Gaussian multiplier bootstrap procedure to approximate $c(\alpha)$. We first generate $N$ normal random vectors of dimension $n \times 1$, denoted as $e_1, e_2, \dots, e_N$. We then calculate pseudo scores $U^{e_b} = \boldsymbol{G}^{\top}\hat{P}^{1/2}e_b/\sqrt{n}$ for $b = 1, \dots, N$. The threshold $c(\alpha)$ is approximated by the $100(1 - \alpha)$-th percentile of pseudo scores:
\[
\hat{c}(\alpha) = f_{\alpha}\left(\norm{U^{e_1}}_{\infty}, \dots, \norm{U^{e_N}}_{\infty}\right),
\]
where $f_{\alpha}(x_1, x_2, \dots, x_N)$ represents the $100(1 - \alpha)$-th percentile of the set $\{x_1, x_2, \dots, x_N\}$.

\subsection{Detection with marginal scores}
\label{subsec:BiRS}
In this subsection, we introduce our signal region detection algorithm combined with the above test statistic $T$. We first introduce some key concepts for signal regions used throughout the paper. Under the alternative hypothesis that there exists signal regions, we define a signal point with index $j$ given that $\beta_j \neq 0$, $1 \leq j \leq p$. Given the polygenic nature of the genome, a genetic region $I$ may contain consecutive signal points. We assume that a signal region satisfies the continuity property if $\beta_I \neq 0$, where the ``$\neq$'' means that no pair of elements is equal. Furthermore, we assume that the signal region $I$ satisfies the separability assumption: for any area that contains $I$, the edges of $I$ are signal points and there are no signal points next to $I$. Lastly, we denote the true signal regions by $I_1^*, \dots, I_\ell^*$ (if exist) and let $\mathcal{I} = \left\{ I_1^*, \dots, I_\ell^* \right\}$.  

Our goal is to determine whether signal regions exist and, if so, to identify their locations. Specifically, we aim first to test: 
\begin{equation}
\label{eq:region global}
H_0: \mathcal{I} = \emptyset \quad\mathrm{v.s.}\quad H_1: \mathcal{I} \neq \emptyset,
\end{equation}
if $H_0$ is rejected, we proceed to detect each signal region in $\mathcal{I}$. Following \cite{zhang2023binary}, the algorithm first compares $T$ with $\hat{c}(\alpha)$ to determine whether signal regions exit. If $T > \hat{c}(\alpha)$, we conduct a binary search procedure that utilizes a sequence of dynamic thresholds generated by the bootstrap vectors to find the specific locations of the signal regions. For an index set $I = \{i_1, \dots, i_r\}$, we define $U(I) = (U_{i_1}, U_{i_2}, \dots, U_{i_r})$ and similar for $U^{e_b}(I)$, $b = 1, \dots, N$. The detailed binary search procedure is as follows.

At the first step, we split the genome into two segments, denoted as $I_{11} = \left\{1, 2, \dots, \lfloor p/2 \rfloor\right\}$ and $I_{12} = \left\{\lfloor p/2 \rfloor + 1, \dots, p\right\}$ with $\lfloor \cdot\rfloor$ taking the value of the largest integer below. Let $I_1 = I_{11}\cup I_{12}$. The test statistic for the $k$th segment is $T_{1k} = \norm{U(I_{1k})}_{\infty}$, for $k = 1, 2$, and the threshold for the two tests is calculated by $\hat{c}_1(\alpha) = f_{\alpha}(\norm{U^{e_1}(I_1)}_{\infty}, \dots, \norm{U^{e_N}(I_1)}_{\infty})$. If $T_{1k} > \hat{c}_1(\alpha)$, then $I_{1k}$ is considered to possibly contain signal regions and is subjected to further binary search procedures. Otherwise, it is concluded that there are no signals within $I_{1k}$. 

Suppose we further conduct the binary search procedure for regions that may contain signals. During the $j$-th binary search, there are $n_j$ segments to be tested ($1 \leq n_j \leq 2^j$), denoted as $I_{j_1}, \dots, I_{jn_{j}}$, with $I_j = \cup_{k=1}^{n_j} I_{j_k}$. The critical value for all $n_{j}$ tests is calculated as $$\hat{c}_j(\alpha) = f_{\alpha}(\norm{U^{e_1}(I_j)}_{\infty}, \dots, \norm{U^{e_N}(I_j)}_{\infty}).$$ If $T_{jk} > \hat{c}_j(\alpha)$ for $k = 1, \dots, n_j$, we perform binary segmentation on $I_{jk}$ to conduct tests in the next iteration of binary search procedure. The segmentation stops if the length of $I_{jk}$ is sufficiently small such that $\abs{I_{jk}} \leq 2^s$, where $s$ is a truncation parameter. This binary search procedure is repeated iteratively until no significant segments remain or all segments become sufficiently small. The detected signal regions are denoted as $\hat{I}_{11}, \dots, \hat{I}_{1l_1}$. We want to emphasis that this binary search step utilizes a sequence of dynamic critical values, which enables the detection of more signals compared to procedures with fixed thresholds. Specifically, since $U^{e_b}(I_j)$ is a sub-vector of $U^{e_b}$, it follows that $\norm{U^{e_b}(I_j)}_{\infty} \leq \norm{U^{e_b}}_{\infty}$, for $b = 1, \dots, N$. This implies $\hat{c}(\alpha) \geq \hat{c}_1(\alpha) \geq \cdots \geq \hat{c}_j(\alpha) \geq \cdots$, which increase the power to detect weaker signals as the binary search progresses.

We next implement a re-search step to detect signals that may have been missed. We substitute the variables within \( \hat{I}_{11}, \dots, \hat{I}_{1l_1} \) with zeros, and repeat the global test and binary search. This process is iterated until the global test confirms that no signals remain or all segments are sufficiently small. We finally rearrange the detected signal regions based on their locations in the genome, denoted as \( \hat{I}_1, \dots, \hat{I}_{\ell} \). The complete detection procedure is referred to as sBiRS and is summarized in Algorithm \ref{alg:sBiRS}.

{\begin{breakablealgorithm}
   \setstretch{1}
    \renewcommand{\algorithmicrequire}{\textbf{Input:}}
    \renewcommand{\algorithmicensure}{\textbf{Output:}}
   \caption{BiRS with summary statistics (sBiRS)}\label{alg:sBiRS}
   \begin{algorithmic}[1]
       \State\textbf{Input}: the vector of marginal scores $U$; the bootstrap vectors $U^{e_1}, \dots, U^{e_N}$; truncation parameter $s$; significant level $\alpha$;
       \hspace*{0.02in}
       \State Calculate $T = \norm{U}_{\infty}$ and $\hat{c}(\alpha) = f_{\alpha}(\norm{U^{e_1}}_{\infty}, \dots, \norm{U^{e_N}}_{\infty})$;
       \If{$T \leq \hat{c}(\alpha)$}
          \State\textbf{Output}: There are no signal regions.
       \Else
          \State Conduct the binary search procedure based on $U$ and $U^{e_1}, \dots, U^{e_N}$ with truncation parameter $s$;
          \State Replace the elements of $U$ and $U^{e_1}, \dots, U^{e_N}$ in estimated signal regions from the last step as zeros;
          \State Recalculate $T$ and $\hat{c}(\alpha)$;
          \While {$T > \hat{c}(\alpha)$}
              \State Conduct the binary search procedure based on the new $U$ and $U^{e_1}, \dots, U^{e_N}$;
              \State Replace the elements of $U$ and $U^{e_1}, \dots, U^{e_N}$ in estimated signal regions from the last step as zeros;
              \State Recalculate $T$ and $\hat{c}(\alpha)$;
          \EndWhile
          \State Rearrange the estimated signal regions to be continuous and separated;
          \State\textbf{Output}: Estimated signal regions $\hat{I}_1, \dots, 
          \hat{I}_{\hat{\ell}}$.
          \EndIf
   \end{algorithmic}
  
\end{breakablealgorithm}}

\subsection{Detection over multiple blocks: a distributed algorithm} 

In whole genome association studies, the number of variants in the genome is about $10^7$. Implementing this searching algorithm requires the use of bootstrap vectors throughout the procedure, which results in excessive memory usage. Additionally, it is impractical to load the entire genotype matrix at once when working with individual-level data. A straightforward solution is to divide the genome into $K$ blocks and detect signal regions within each block using a significant level of $\alpha/K$ (i.e., the Bonferroni correction), but this approach is conservative in most cases. Alternatively, detection procedures can be performed with fixed thresholds by applying global thresholds to control the detection results within each block. While this method avoids conservativeness, it reduces the power of sBiRS due to the use of higher critical values compared to dynamic thresholds which are adjusted based on prior testing results. This issue is analogous to the power reduction phenomenon observed in distributed learning \citep{cai2022distributed}, where large-scale learning tasks are performed across multiple nodes. To address this problem, we introduce a distributed version of our detection algorithm in this subsection. This approach preserves the advantages of dynamic thresholds while maintaining control over family-wise error rates (FWERs) and false discovery rates (FDRs).

We divide the entire region into \( K \) blocks, denoted as \( B_1, \dots, B_K \). For block \( k \) (\( k = 1, \dots, K \)), the vector of marginal scores is represented by \( U(B_k) \), and the corresponding bootstrap vectors are \( U^{e_1}(B_k), \dots, U^{e_N}(B_k) \). Within each block, a local machine applies the BiRS algorithm, resulting in a collection of detected signal regions $
\hat{\mathcal{I}}_k = \left\{ \hat{I}_{1}^{(k)}, \dots, \hat{I}_{\ell_k}^{(k)} \right\}$ for block $k$, $k = 1, \dots, K$. The union of the detected signal regions in block \( k \) is denoted as \( \hat{I}^{(k)} = \cup_{i=1}^{\ell_k} \hat{I}_{i}^{(k)} \). The corresponding collection of maximum marginal scores is defined as:
\[
\mathcal{T}_k = \left\{T(B_k), T(\hat{I}_{1}^{(k)}), \dots, T(\hat{I}_{\ell_k}^{(k)}) \right\} = \left\{\|U(B_k)\|_{\infty}, \|U(\hat{I}_{1}^{(k)})\|_{\infty}, \dots, \|U(\hat{I}_{\ell_k}^{(k)})\|_{\infty} \right\}.
\] For variables related to the bootstrap vectors, we define the following quantities for block \( k \):
\[
\begin{aligned}
    M(B_k) & = \left(M_1(B_k), \dots, M_N(B_k)\right)^\top = \left(\|U^{e_1}(B_k)\|_{\infty}, \dots, \|U^{e_N}(B_k)\|_{\infty}\right)^\top, \\
    L(B_k) & = \left(L_1(B_k), \dots, L_N(B_k)\right)^\top = \left(\|U^{e_1}(\hat{I}^{(k)})\|_{\infty}, \dots, \|U^{e_N}(\hat{I}^{(k)})\|_{\infty}\right)^\top.
\end{aligned}
\] Finally, we transfer the block results 
\[
\mathcal{R}_k = \left\{ \hat{\mathcal{I}}_k, \mathcal{T}_k, M(B_k), L(B_k) \right\}, \quad k = 1, \dots, K,
\] to the central machine for further analysis.

In the central machine, we treat these blocks as potential signal points and apply the sBiRS procedure to identify which blocks are significant.  Precisely, we define 
$$\tilde{U} = (T(B_1), \dots, T(B_K))^\top,$$
where each component represents the maximum marginal score within each block. The corresponding bootstrap vectors for $b$-th boostrap are defined as $$\tilde{U}^{e_b} = (M_b(B_1), \dots, M_b(B_K))^\top, b = 1, \dots N.$$ We run the sBiRS algorithm with $\tilde{U}, \tilde{U}^{e_1}, \dots, \tilde{U}^{e_N}$, truncation parameter $s = 0$ and significant level $\alpha$. Suppose the significant blocks identified by sBiRS are $B_{j_1}, \dots, B_{j_{\hat{k}}}$, where $j_1, \ldots, j_{\hat{k}}$ are indices of significant blocks. We remove insignificant blocks, even if there are signal regions detected by the local machine, and then perform a control procedure for the detected signal regions within the significant blocks. For the $b$-th boostrap vector, we define $$\tilde{L}^{e_b} = \left(L_b(B_{j_1}), \dots, L_b(B_{j_{\hat{k}}})\right)^\top=\left(\|U^{e_b}(\hat{I}^{(j_{1})})\|_{\infty}, \dots, \|U^{e_b}(\hat{I}^{(j_{\hat{k}})})\|_{\infty}\right)^\top.$$ 
Based on the \( \tilde{L}^{e_b} \)'s, we calculate new critical values \( \hat{c}_{\min}(\alpha) \) for the test statistics in each signal region within the significant blocks:
 $$\hat{c}_{\min}(\alpha) = f_{\alpha}(\tilde{L}^{e_1}, \dots, \tilde{L}^{e_N}).$$ For the $r$-th block and $i$-th signal region within the block ($r = j_{1}, \dots, j_{\hat{k}}$, $i = 1, \dots, \ell_r$), the signal region $\hat{I}_i^{(r)}$ is significant if $T_n(\hat{I}_i^{(r)}) > \hat{c}_{\min}(\alpha)$. This distributed algorithm is summarized in Algorithm \ref{alg:dBiRS}. 

Since the significance level for the detection procedure within each block was set to \(\alpha\), the family-wise error rate (FWER) across multiple tests on the final estimated signal regions may not be controlled. Therefore, it is necessary to construct the threshold \(\hat{c}_{\min}(\alpha)\) to ensure FWER control by filtering out less significant signal regions. This control procedure also helps manage the proportion of false discoveries, leading to consistent detection results, as demonstrated in the proof of Theorem \ref{thm:FDR} in the Supplementary Material.

{\begin{breakablealgorithm}
   \setstretch{1}
    \renewcommand{\algorithmicrequire}{\textbf{Input:}}
    \renewcommand{\algorithmicensure}{\textbf{Output:}}
   \caption{Distributed BiRS (dBiRS)}\label{alg:dBiRS}
   \begin{algorithmic}[1]
       \State\textbf{Input}: the vector of marginal scores $U$; the bootstrap vectors $U^{e_1}, \dots, U^{e_N}$; truncation parameter $s$; significant level $\alpha$; number of blocks $K$;
       \hspace*{0.02in}
       \State Divide the genome into $K$ blocks $B_1, \dots, B_K$;
       \For{$k = 1, \dots, K$}
           \State Run $\mathrm{sBiRS}(U(B_k), U^{e_1}(B_k), \dots, U^{e_N}(B_k), s)$ to derive the collection of detection results $\mathcal{R}_k$ in block $B_k$;
       \EndFor
       \State Let $\tilde{U} = (T(B_1), \dots, T(B_K))^\top$ and $\tilde{U}^{e_b} = (M_b(B_1), \dots, M_b(B_K))^\top$, $b = 1, \dots, N$; 
       \State Run $\mathrm{sBiRS}(\tilde{U}, \tilde{U}^{e_1}, \dots, \tilde{U}^{e_N}, 0)$ to get the significant blocks $j_1, \dots, j_{\hat{k}}$, $\hat{k} \leq K$; 
       \State Let $\hat{c}_{\min}(\alpha) = f_{\alpha}(\tilde{L}^{e_1}, \dots, \tilde{L}^{e_N})$, where $\tilde{L}^{e_b} = (L_b(B_{j_1}), \dots, L_b(B_{j_{\hat{k}}})^\top$;
       \For{$r = j_1, j_2, \dots j_{\hat{k}}$}
           \For{$i = 1, \dots, \ell_{r}$}
               \If{$T_n(\hat{I}_i^{(r)}) > \hat{c}_{\min}(\alpha)$}
                  \State Take $\hat{I}_i^{(r)}$ as one of the estimated signal regions;
                \EndIf
            \EndFor
        \EndFor
       \State Rearrange these estimated signal regions to be continuous and separable;
       \State\textbf{Output}: Estimated signal regions $\hat{I}_1, \dots, \hat{I}_{\hat{\ell}}$.
   \end{algorithmic}
\end{breakablealgorithm}}

In practice, we typically select blocks with dimensions between \(10^3\) and \(10^4\). Consequently, the number of blocks \(K\) is approximately $10^3$ to $10^4$, given that the total dimension \(p\) may exceed \(10^7\). Because signal regions are usually sparsely distributed across the genome, most blocks are unlikely to contain signals.
Applying sBiRS to determine the significance of blocks will sequentially remove most of the
blocks and result in a reduction of dynamic critical values, which helps find more significant blocks with relatively weaker signals. Moreover, this procedure is equivalent to running sBiRS with truncation $
\log_2(\max_k p_k) \leq s \leq \log_2(\min_k p_k) + 1$, where \(p_k\) represents the dimension of block \(B_k\) for \(k = 1, \dots, K\).  

\subsection{Theoretical guarantees} 
For the flow of exposition, we provide a description of the theoretical results here, and defer the technical presentation of the family-wise error rate and the detection accuracy in the Appendix. For size analysis, after assuming some mild conditions to the exponential family, the dBiRS algorithm asymptotically controls the family-wise error rate of test problem \eqref{eq:region global}. The method allows the dimension to grow at an exponential rate relative to the sample size. For detection accuracy analysis, we prove that the dBiRS algorithm overcomes the power reduction phenomenon and maintains the facilitated properties of the BiRS algorithm under GLM model. Specifically, when there are model misspecifications under the alternative hypothesis, the dBiRS algorithm can still achieve detection consistency, even when the signal strength of the signal regions decays at a certain rate. In contrast, the Q-SCAN method requires balanced signal strengths across regions. Moreover, the dBiRS algorithm relaxes the M-dependence assumption to a "weak" dependence assumption, which permits long-range correlation (LD) and is more suitable to genetic association studies. Additionally, with an appropriate block-splitting strategy, the dBiRS algorithm imposes fewer restrictions on the lengths of signal regions, which enables consistent detection of both shorter or longer signal regions compared to Q-SCAN. See Theorem \ref{thm:FWER}, \ref{thm:TPR} and \ref{thm:FDR} for more details.

\section{Simulation Study}
\label{sec:simulation}
We conduct simulation studies to compare the proposed dBiRS method with the state-of-art Q-SCAN and KnockoffScreen procedure. We generate sequence data of European ancestry from 10,000 chromosomes across 5-megabase (Mb) regions using the calibrated coalescent model \citep{cosi2014}, and the total number of variants is 349,640. We evaluate the family-wise error rate (FWER), false discovery rate (FDR), signal region detection rate (DR), and true positive rate (TPR) for both continuous and binary phenotypes.

For the evaluation of FWER, the continuous phenotypes are generated using the model:
\begin{equation*}
    Y = 0.5X_1 + 0.5X_2 + \varepsilon,
\end{equation*}
where $X_1$ is a continuous covariate sampled from a standard normal distribution, $X_1 \sim \mathcal{N}(0,1)$, and $X_2$ is a dichotomous covariate that takes values $0$ and $1$ with equal probability. The random noise $\varepsilon$ is generated from a standard normal distribution, $\varepsilon \sim \mathcal{N}(0,1)$. The dichotomous phenotypes are generated using the following logistic regression model:
\begin{equation*}
    \mathrm{logit}\left\{ \mathbb{P}(Y = 1) \right\} = 0.5X_1 + 0.5X_2,
\end{equation*}
where $\mathbb{P}(Y = 1) = (\mathbb{P}(Y_1 = 1), \dots, \mathbb{P}(Y_n = 1))^\top$. We perform dBiRS and Q-SCAN analyses based on 1,000 Monte Carlo runs under the linear and logistic models to compute FWERs. For both dBiRS and Q-SCAN, the number of bootstrap iterations is set to 1,000. We do not calculate the FWER for KnockoffScreen because it only controls the FDR \citep{he2021identification}. The empirical FWERs estimated for dBiRS and Q-SCAN are presented in Table \ref{table:FWER} for significance levels $\alpha = 0.01$ and $\alpha = 0.05$. The FWERs of Q-SCAN and dBiRS are accurate at both significance levels, demonstrating that both methods effectively control the FWER.

\begin{table}[h]
  \setlength{\abovecaptionskip}{0.1cm}
  \caption{FWERs of dBiRS and Q-SCAN in continuous and dichotomous phenotypes \label{table:FWER}} 
  \setlength{\tabcolsep}{10mm}
  \renewcommand\arraystretch{0.6}
  \centering
    \begin{tabular}{ccccc}
    \toprule
    & \multicolumn{2}{c}{continuous} & \multicolumn{2}{c}{dichotomous}  \\
    \cmidrule(r){2-3} \cmidrule(r){4-5} 
    $\alpha$ & 0.01 & 0.05 & 0.01 & 0.05  \\
    \midrule
    dBiRS & 0.011  & 0.053 & 0.011 & 0.051  \\
    Q-SCAN & 0.011  & 0.051 & 0.009 & 0.048  \\
    \bottomrule
    \end{tabular}%
\end{table}

We next evaluate the detection accuracy of dBiRS and compare its performance with Q-SCAN and KnockoffScreen. We randomly select four 5kb causal windows, denoted as $I_i$, $i = 1, \dots, 4$.  We generate continuous and dichotomous phenotypes by
\begin{equation*}
    Y = 0.5X_1 + 0.5X_2 + \boldsymbol{G}_{I_1}\beta_{I_1} + \cdots + \boldsymbol{G}_{I_4}\beta_{I_4} + \varepsilon,
\end{equation*}
and 
\begin{equation*}
    \mathrm{logit}\left\{ \mathbb{P}(Y = 1) \right\} = 0.5X_1 + 0.5X_2 + \boldsymbol{G}_{I_1}\beta_{I_1} + \cdots + \boldsymbol{G}_{I_4}\beta_{I_4},
\end{equation*}
where $\boldsymbol{G}_{I_1}, \dots, \boldsymbol{G}_{I_4}$ represent the genotypes of the causal windows, and $\beta_{I_1}, \dots, \beta_{I_4}$ are the corresponding effect sizes. In each causal window, $10\%$ of the variants are randomly designated as causal, and each causal variant is assigned an effect size as a decreasing function of the minor allele frequency (MAF), i.e., $\abs{\beta} = c\abs{\log_{10}(\mathrm{MAF})}$. The parameter $c$ is set to $c \in \left\{ 0.12, 0.15 \right\}$ for continuous outcomes and $c \in \left\{ 0.25, 0.30 \right\}$ for dichotomous outcomes. The sign of $\beta$ is randomly assigned with $50\%$ of the values being positive and $50\%$ negative. 

We evaluate the DR,TPR and $h$ kilobase (kb) FDR for three methods based on 100 Monte Carlo runs across different values of $c$ \citep{li2022simultaneous,he2021identification}. Specifically, we denote the true signal regions as $I_1^*, \dots, I^*_{\ell}$ and the estimated signal regions as $\hat{I}_1, \dots, \hat{I}_{\hat{\ell}}$. Let $I^* = \cup_{i=1}^{\ell} I^*_i$ represent the union of the true signal regions, and $\hat{I} = \cup_{i=1}^{\hat{\ell}} \hat{I}_i$ represent the union of the estimated signal regions.  The DR and TPR are defined as 
\begin{equation*}
    \mathrm{DR} = \frac{1}{\ell}\mathbb{E}\sum_{i=1}^{\ell} \boldsymbol{1}\left\{ \hat{I} \cap I^*_{\ell} \neq \emptyset \right\}, \quad \mathrm{TPR} = \mathbb{E}\left(\abs{I^*\cap \hat{I}}/\abs{I^*}\right). 
\end{equation*}
Following \cite{he2021identification}, we define the FDR($h$) for $h \in \{ 25,50,75 \}$ as 
\begin{equation*}
    \mathrm{FDR}(h) = \mathbb{E}\left(\abs{\left\{ j: j \in \hat{I}, d(j, \hat{I}) \geq h \right\}}/\abs{\hat{I}}\right)
\end{equation*}
to account for the spill-over effect of the LD structure, where $d(j, I^*) = \min_{i \in I^*} \abs{i - j}$ is the minimum distance between point $j$ and true signal regions. The DR represents the proportion of true signal regions that are detected and measures the ability to identify signal regions. In contrast, the TPR and FDR($h$) evaluate how well the true signal regions are recovered by assessing the similarity or distance between the true and identified signal regions. The DR, TPR, and FDR($h$) of the three methods for continuous and dichotomous phenotypes are presented in Table \ref{table:mean_results}. Additionally, we provide the standard deviations (SDs) for DR, TPR, and FDR($h$) in Table \ref{table:sd_results}.

\begin{table}[h]
  \setlength{\abovecaptionskip}{0.1cm}
  \caption{Detection results (DR, TPR, FDR($h$)) for continuous and dichotomous phenotypes, $h = 25, 50, 75$. \label{table:mean_results}} 
  \setlength{\tabcolsep}{5.5mm}
  \renewcommand\arraystretch{0.6}
  \centering
    \begin{tabular}{cccccc}
    \toprule
     & DR & TPR & FDR(25) & FDR(50) & FDR(75) \\
    \midrule
    \multicolumn{6}{c}{continuous, $c = 0.12$}\\
    dBiRS & $\boldsymbol{0.990}$  & $\boldsymbol{0.830}$  & $\boldsymbol{0.192}$  & $\boldsymbol{0.049}$  & $\boldsymbol{0.023}$ \\
    Q-SCAN & 0.980  & 0.783  & 0.234  & 0.077  & 0.034 \\
    KnockoffScreen & 0.980  & 0.227  & 0.475  & 0.379  & 0.287 \\
    \midrule
    \multicolumn{6}{c}{continuous, $c = 0.15$}\\
     dBiRS  & $\boldsymbol{1.000}$  & $\boldsymbol{0.881}$  & $\boldsymbol{0.285}$  & $\boldsymbol{0.094}$ & $\boldsymbol{0.037}$ \\
    Q-SCAN  & $\boldsymbol{1.000}$  & 0.837  & 0.326  & 0.113 & 0.040 \\
    KnockoffScreen  & $\boldsymbol{1.000}$  & 0.505  & 0.605  & 0.516  & 0.408 \\
    \midrule
    \multicolumn{6}{c}{dichotomous, $c = 0.25$}\\
    dBiRS & $\boldsymbol{0.990}$  & $\boldsymbol{0.793}$  & $\boldsymbol{0.183}$  & $\boldsymbol{0.049}$  & $\boldsymbol{0.020}$ \\
    Q-SCAN & 0.930  & 0.682  & 0.203  & 0.077  & 0.039 \\
    KnockoffScreen & 0.960  & 0.184  & 0.495  & 0.416  & 0.237 \\
    \midrule
    \multicolumn{6}{c}{dichotomous, $c = 0.30$}\\
     dBiRS  & $\boldsymbol{1.000}$  & $\boldsymbol{0.853}$  & $\boldsymbol{0.235}$  & $\boldsymbol{0.058}$ & $\boldsymbol{0.019}$ \\
    Q-SCAN  & 0.980  & 0.766  & 0.277  & 0.082 & 0.041 \\
    KnockoffScreen  & 0.990  & 0.361  & 0.533  & 0.440  & 0.294 \\
    \bottomrule
    \end{tabular}%
\end{table}

\begin{table}[h]
  \setlength{\abovecaptionskip}{0.1cm}
  \caption{Standard deviations for DR, TPR, FDR($h$) in continuous and dichotomous phenotypes, $h = 25, 50, 75$. \label{table:sd_results}} 
  \setlength{\tabcolsep}{3mm}
  \renewcommand\arraystretch{0.6}
  \centering
    \begin{tabular}{cccccc}
    \toprule
     & sd(DR) & sd(TPR) & sd(FDR(25)) & sd(FDR(50)) & sd(FDR(75)) \\
    \midrule
    \multicolumn{6}{c}{continuous, $c = 0.12$}\\
    dBiRS & $\boldsymbol{0.050}$  & $\boldsymbol{0.065}$  & $\boldsymbol{0.051}$  & $\boldsymbol{0.034}$  & $\boldsymbol{0.027}$ \\
    Q-SCAN & 0.069  & 0.105  & 0.083  & 0.056  & 0.052 \\
    KnockoffScreen & 0.069  & 0.178  & 0.254  & 0.249  & 0.203 \\
    \midrule
    \multicolumn{6}{c}{continuous, $c = 0.15$}\\
     dBiRS  & $\boldsymbol{0.000}$  & $\boldsymbol{0.037}$  & $\boldsymbol{0.046}$  & $\boldsymbol{0.039}$ & $\boldsymbol{0.031}$ \\
    Q-SCAN  & $\boldsymbol{0.000}$  & 0.072  & 0.073  & 0.050 & 0.042 \\
    KnockoffScreen  & $\boldsymbol{0.000}$  & 0.220  & 0.111  & 0.091  & 0.074 \\
    \midrule
    \multicolumn{6}{c}{dichotomous, $c = 0.25$}\\
    dBiRS & $\boldsymbol{0.05}$  & $\boldsymbol{0.07}$  & $\boldsymbol{0.088}$  & $\boldsymbol{0.040}$  & $\boldsymbol{0.031}$ \\
    Q-SCAN & 0.115  & 0.126  & 0.096  & 0.066  & 0.050 \\
    KnockoffScreen & 0.093  & 0.119  & 0.210  & 0.214  & 0.188 \\
    \midrule
    \multicolumn{6}{c}{dichotomous, $c = 0.30$}\\
     dBiRS  & $\boldsymbol{0.000}$  & $\boldsymbol{0.039}$  & $\boldsymbol{0.079}$  & $\boldsymbol{0.037}$ & $\boldsymbol{0.024}$ \\
    Q-SCAN  & 0.069  & 0.102  & 0.107  & 0.068 & 0.044 \\
    KnockoffScreen  & 0.050  & 0.168  & 0.131  & 0.121  & 0.123 \\
    \bottomrule
    \end{tabular}%
\end{table}

Table \ref{table:mean_results} shows that dBiRS achieves the highest DRs and TPRs with the lowest FDRs across all signal strengths $c$ and for all values of the distance parameter $h = 25, 50, 75$ in both continuous and dichotomous phenotypes. As the value of $h$ increases, the FDR of all methods decreases, and the FDR of dBiRS and Q-SCAN falls below 0.05 when $h = 75$ kb. Q-SCAN outperforms KnockoffScreen in recovering true signal regions by achieving higher TPR and lower FDR. However, it identifies a smaller proportion of true signal regions (i.e., a lower DR) compared to KnockoffScreen. KnockoffScreen employs LD-pruning with a 0.75 correlation threshold to remove highly correlated variants before conducting the analysis. It identifies a larger number of signal regions, which increases the chance of overlap with true signal regions and thereby boosting the DR. However, the signal regions identified by KnockoffScreen have lower coverage of true signal regions due to both the LD-pruning process and a higher number of false discoveries. Additionally, Table \ref{table:sd_results} shows that dBiRS has the lowest standard deviations in all cases, indicating that dBiRS is the most stable method.

\begin{figure}[htbp]
\centering
\includegraphics[width=15cm]{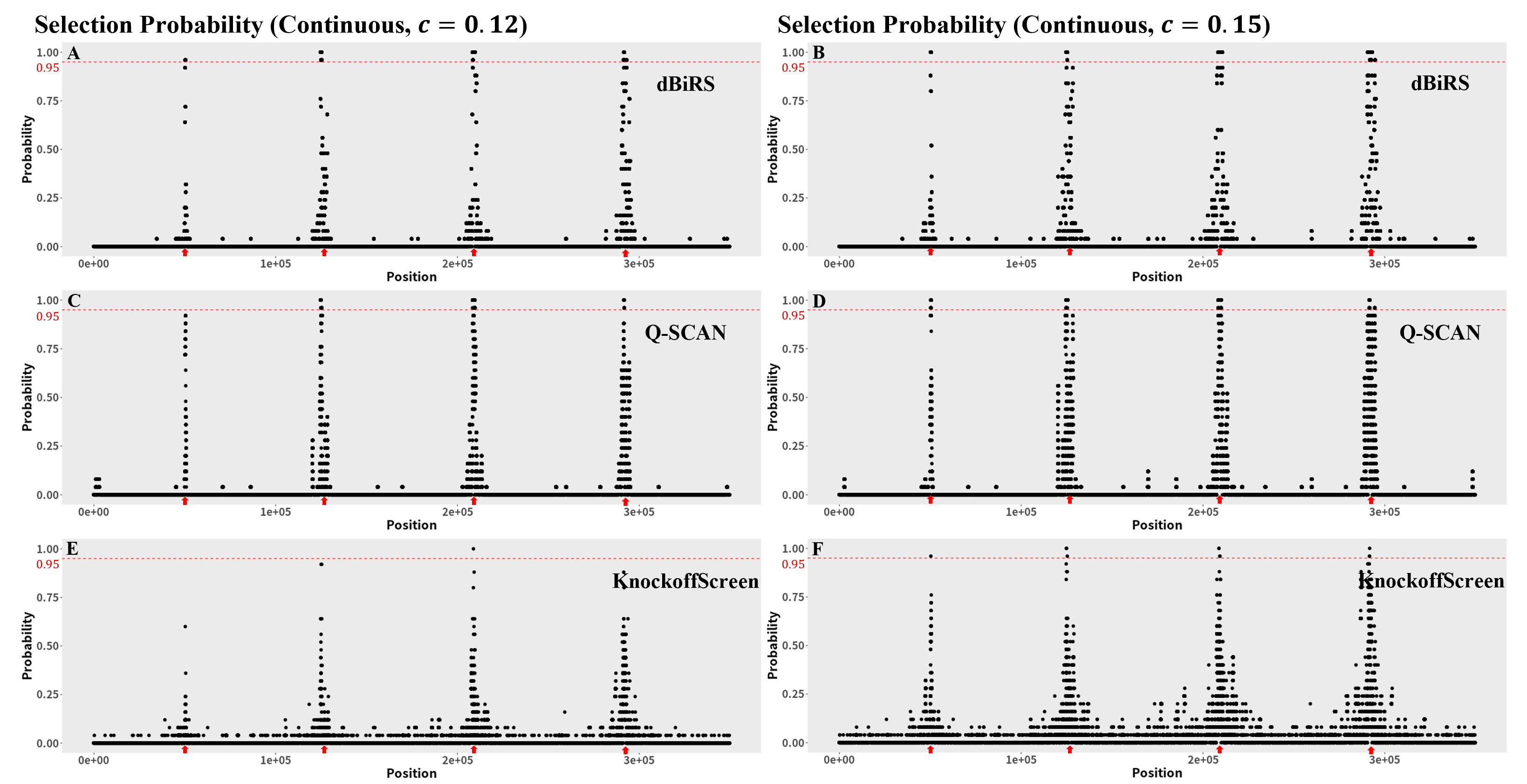}
\caption{\small The selection probabilities of dBiRS, Q-SCAN and KnockoffScreen under dichotomous phenotype. The left panel is the selection probabilities when $c = 0.25$ and the right panel is the selection probabilities when $c = 0.30$. The red arrows refer to the locations of four true signal regions.}
\label{fig:variational_continuous}
\end{figure}

\begin{figure}[htbp]
\centering
\includegraphics[width=15cm]{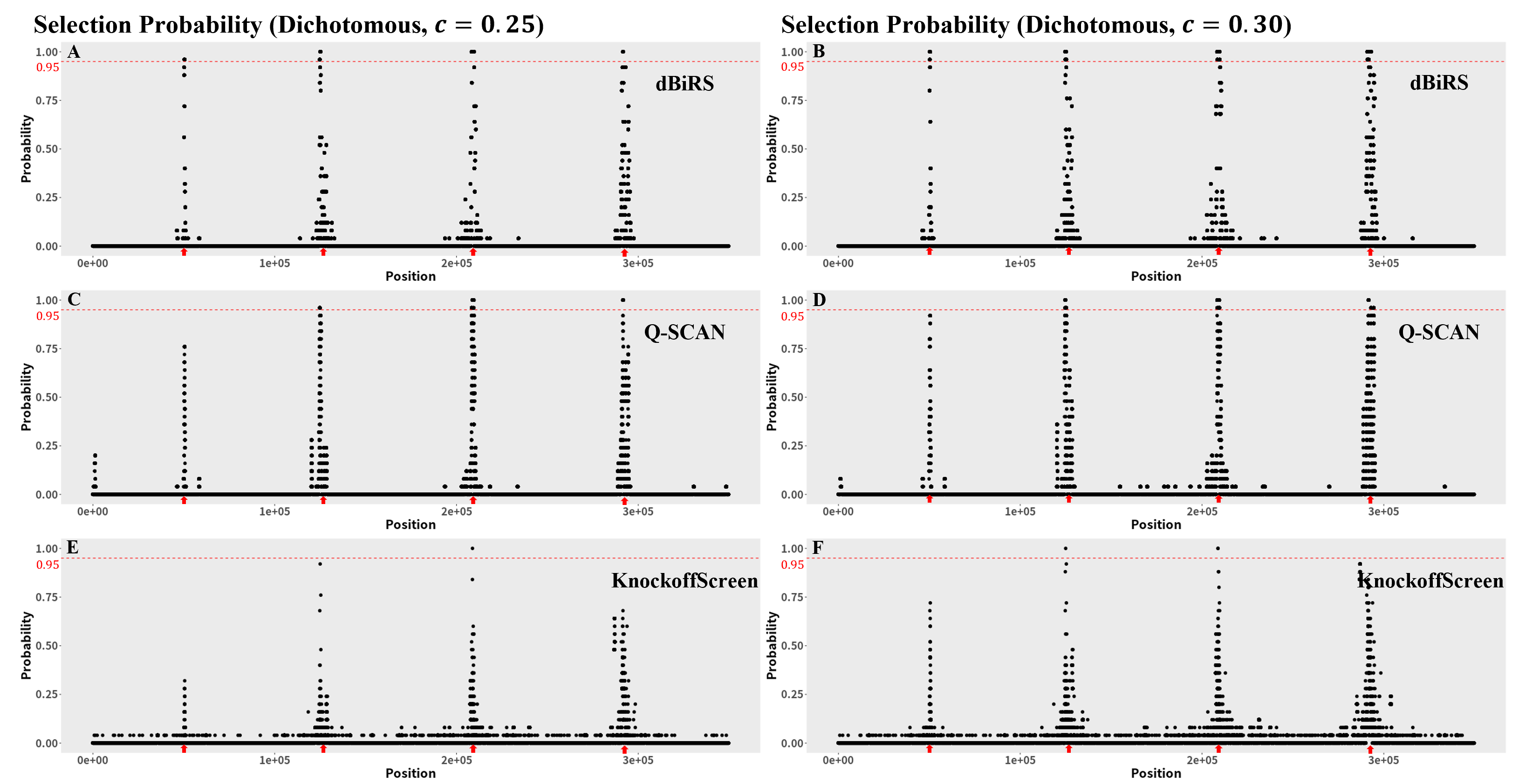}
\caption{\small The selection probabilities of dBiRS, Q-SCAN and KnockoffScreen under dichotomous phenotype. The left panel is the selection probabilities when $c = 0.25$ and the right panel is the selection probabilities when $c = 0.30$. The red arrows refer to the locations of four true signal regions.}
\label{fig:variational_binary}
\end{figure}

To further illustrate the recovery of true signal regions across different methods, we present the selection probabilities for all variants, where the selection probability of a variant $j$ is defined as the proportion of replications in which it is identified as a signal variant.  We present the selection probabilities under the continuous phenotype and the dichotomous phenotype in Figure \ref{fig:variational_continuous} and Figure \ref{fig:variational_binary}, respectively.

While all procedures exhibit relatively high selection probabilities around the four true signal regions, dBiRS demonstrates the greatest accuracy and stability, achieving selection probabilities exceeding 0.95 in all true signal regions across all settings. When signals are relatively weak (i.e., $c = 0.12$ for the continuous trait in Figure 1A and $c = 0.25$ for the binary trait in Figure 2A), Q-SCAN shows relatively low selection probabilities in the leftmost true signal region (i.e., selection probability $< 0.95$) and identifies some variants in non-signal regions at the beginning of the genome. This suggests that the Q-SCAN procedure is more prone to producing misleading results across replications, which is consistent with results that show a higher FDR compared to dBiRS. KnockoffScreen demonstrates the lowest selection probabilities in the four true signal regions and identifies only one variant with a selection probability greater than 0.95 (Figure 1A and Figure 2A). When the signal is stronger (i.e., $c = 0.15$ for the continuous trait in Figure 1B and $c = 0.30$ for the binary trait in Figure 2B), KnockoffScreen exhibits substantially higher selection probabilities in non-signal regions compared to both Q-SCAN and dBiRS (Figure 1B). The spread of selection probabilities across non-signal regions highlights a lack of stability for KnockoffScreen and indicates its tendency to identify false positives for spurious variants. Based on these results, we conclude that the dBiRS method outperforms the other procedures in terms of both detection accuracy and robustness across different settings.

\section{Application}
\label{sec:application}
Extensive research has identified common genetic variants associated with cognitive health, but the roles of rare genetic variants, which often have more subtle and complex effects, remain poorly understood. In this study, we apply the dBiRS algorithm to conduct Whole Exome Sequencing (WES) analyses on the core cognitive phenotypes: fluid intelligence (Field ID 20016) and prospective memory (Field ID 20018), using data from the UK Biobank (\url{https://biobank.ctsu.ox.ac.uk/crystal/index.cgi}). The WES data used in this study is derived from the final exome release in PLINK format (Field ID 23158) from the UK Biobank. Fluid intelligence reflects problem-solving and reasoning abilities, while prospective memory measures the capacity to remember and execute planned actions, a critical daily-life function that often declines with age.  These phenotypes are closely linked to aging and neurodegeneration. By analyzing these traits, we aim to uncover rare variants that may confer either risk or resilience to cognitive functions, providing insights into the genetic and biological mechanisms underlying cognitive impairment. This approach has the potential to deepen our understanding of cognitive aging and uncover genetic factors contributing to the risk of neurodegenerative disorders such as Alzheimer’s disease (AD).
 
We implemented a standard quality control procedure to ensure the integrity of the dataset before analysis. First, we excluded individuals flagged as outliers by the UK Biobank based on genotyping missingness rates or heterogeneity, as well as those whose genotypically inferred sex did not align with their self-reported sex. To address population stratification, we utilized principal component analysis provided by the UK Biobank and excluded individuals identified as non-European. Specifically, we removed individuals whose values for either of the first two principal components deviated by more than five standard deviations from the mean. We also excluded participants who self-reported an ethnicity other than European. Furthermore, individuals with more than $5\%$ missing genotype data across variants passing UK Biobank's quality control were removed. For variant-level quality control, we retained only biallelic autosomal variants assayed by both genotyping arrays used by the UK Biobank. Variants that failed UK Biobank quality control in any genotyping batch were excluded. Additionally, we removed variants with a Hardy-Weinberg equilibrium (HWE) $p$-value below $10^{-50}$ or with a minor allele count (MAC) $\leq 1$. This stringent filtering resulted in 13,681,006 variants across 22 chromosomes in the WES dataset, 154,785 samples for fluid intelligence analysis, and 155,448 samples for prospective memory analysis. After the quality control procedure, the dBiRS analysis was performed using summary statistics derived from a generalized linear model. The model adjusted for covariates, including sex, age, assessment center, and the top five genomic principal components provided by the UK Biobank.

Figure 1 illustrates the distribution of functional consequences of variants identified by dBiRS, with a substantial number of variants located in exonic, intronic, and UTR regions. Variants in the exonic region may directly alter protein structure and function. Variants in the 3' UTR (UTR3) are likely involved in post-transcriptional regulation, including mRNA stability and microRNA binding. Intronic variants could disrupt splicing mechanisms or long-range regulatory elements, further impacting gene function. While exonic variants are commonly the focus of WES studies, our findings also underscore the important role of rare variants in non-coding and regulatory regions, such as UTRs and intronic regions, which have historically been understudied in the context of intelligence and prospective memory.

For the study on intelligence, dBiRS identified signal regions encompassing 327 genes, including 84 genome-wide significant genes previously reported for their association with intelligence. Additionally, there are 11 genome-wide significant genes overlapping with those associated with general cognitive function, 30 genes overlapping with those implicated in schizophrenia, 65 genes associated with educational attainment, and 35 genes overlapping with autism spectrum disorder. Based on the GWAS Catalog, 33 genome-wide significant common variants were also identified in our WES study. Notable examples include SNP rs2271928 in COL16A1, SNP rs11264680 in CRTC2, and SNP rs539096 in PTPRF, among others. 

The dBiRS method identifies numerous additional rare variant associations in the exonic, intronic, and 3' UTR regions of genes that are previously reported to be associated with intelligence or cognitive performance. For example, a previous study highlights the significance of CRTC2 in endothelial function, which is essential for maintaining vascular physiology in the brain \citep{kanki2020creb}. While prior genome-wide studies have identified common variants in CRTC2 associated with cognitive traits \citep{savage2018genome}, the dBiRS method uncovers 4 rare variants in the exonic region of CRTC2, which may directly alter protein function, 3 rare variants in the intronic region that could influence gene regulation, and 4 rare variants in the 3' UTR region, potentially affecting mRNA stability and translation efficiency. Another example is BRWD1 that is previously identified as a significant gene associated with intelligence \citep{savage2018genome}. In our study, we identified rare variants in its exonic, intronic, and 3' UTR regions. Research has shown that BRWD1 plays a role in establishing epigenetic states during B cell development, which is essential for proper immune function \citep{mandal2018brwd1}. Further studies are needed to clarify its specific contributions to neural development and cognition.
\begin{figure}[htbp]
\centering
\includegraphics[width=14cm]{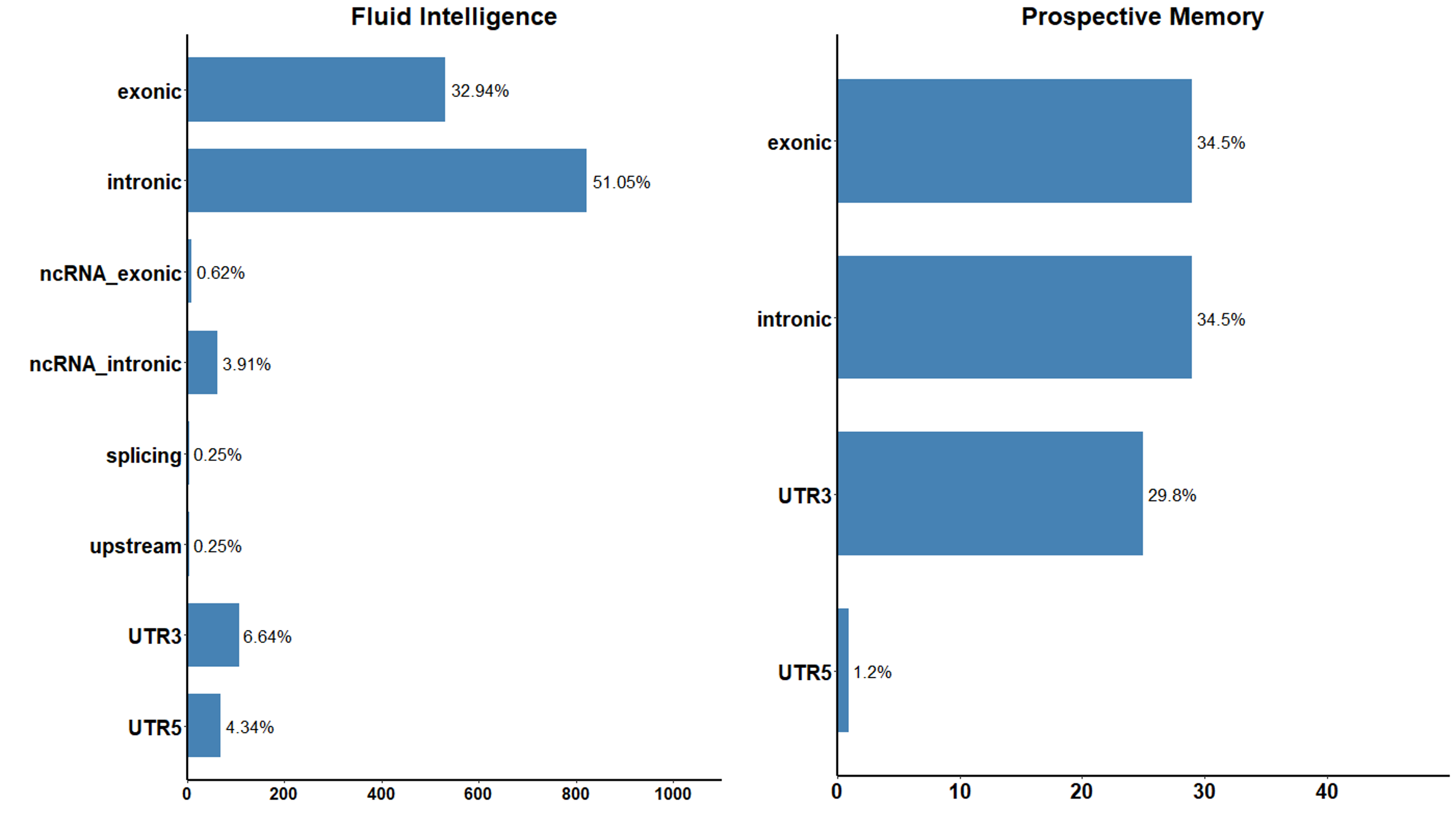}
\caption{\small Distribution of the functional consequences of SNPs in signal regions identified by dBiRS.}
\label{fig:application}
\end{figure}

The dBiRS method identified rare variants in 22 novel genes that have not been previously reported in the GWAS Catalog for intelligence. Among these 22 genes, 14 were significant in the gene-based test for general cognitive ability \citep{davies2018study}. Of the remaining 9 genes, SZT2 has been linked to common variants associated with Alzheimer’s disease and specific brain regions \citep{herold2016family}; ITIH4 has been associated with schizophrenia \citep{ripke2011schizophrenia}; and NDUFA6 has been linked to brain connectivity measurements \citep{zhang2023brain}. Genes SLC39A8, TOP2B, NKIRAS1, GLTBD1, and TPM3 are new findings that have not been previously reported in relation to cognitive performance measurements. Among these genes, TOP2B plays a critical role in neuronal development by regulating gene expression during brain development \citep{tiwari2012target}. Proper functioning of TOP2B is essential for neuronal connectivity and synaptic plasticity, both of which underlie learning and memory \citep{madabhushi2015activity}. Disruptions in TOP2B have been linked to neurodevelopmental disorders, which could implicate it in intelligence \citep{king2013topoisomerases}.

For the study on prospective memory, we identified eight novel genes associated with prospective memory, a key component of cognitive function. Among these, OMA1 stands out due to its established links with multiple neuroimaging measurements, including white matter lesion progression and brain column structure, underscoring its potential role in cognitive processes \citep{wang2020genome}. Notably, our analysis uncovered 10 rare variants in the 3' UTR, 6 rare variants in the exonic region, 7 rare variants in the intronic region, and 1 common variant in the 5' UTR of OMA1. These findings suggest that OMA1 may influence prospective memory through diverse mechanisms, such as protein function modulation, gene regulation, and mRNA stability. Similarly, we identified variants in the intronic, exonic, and 3' UTR regions of CNGB3, further emphasizing the importance of regulatory regions in prospective memory. CNGB3 is known to be associated with educational attainment \citep{okbay2022polygenic}, cognitive function \citep{lee2018gene} and anxiety \citep{meier2019genetic}, which supports its potential role in cognitive traits. Additionally, we discovered HIGD1B, previously reported to be associated with educational attainment \citep{pasman2022genetic}, and SFMBT2 which has been linked to anxiety and stress-related disorders \citep{hollins2016microrna}. Other novel findings include the genes NT5C3B, NOSTRIN, and APOD, which represent unexplored candidates for prospective memory and related cognitive traits. These discoveries underscore the value of our approach in uncovering novel genetic mechanisms underlying complex cognitive traits and highlight new directions for future research into the genetic architecture of prospective memory.

\section{Discussion}
\label{sec:discussion}
We developed a computationally efficient distributed Binary and Re-Search (dBiRS) algorithm by incorporating the infinity norm of regression-based summary statistics to support the analysis of both binary and continuous outcomes for whole-genome and whole-exome sequencing studies. Compared to scan-based algorithms and the knockoffScreen method, dBiRS demonstrates greater power while maintaining strict control over family-wise error rates (FWER) and false discovery rates (FDR). Furthermore, dBiRS enables parallel computing of block-wise results, which are then aggregated through a central machine to ensure both detection accuracy and computational efficiency. Empirical studies further demonstrate its robustness and adaptability, even under conditions of signal decay.

Further extensions of dBiR could explore the integration of functional annotations. While the current framework effectively detects signal regions without annotations, their incorporation may provide additional biological context and improve detection accuracy for rare or weak signals in less characterized regions of the genome. Our framework is also flexible to incorporate various functional annotations by directly adding weights into SNP-level summary statistics, where weights can be estimated using annotation principal components \citep{zhou2023favor}, CADD \citep{kircher2014general} or other methods. Additionally, further validation of the identified associations through experimental studies is necessary to elucidate their functional relevance and causal mechanisms.

\section*{Appendix: Theoretical Guarantees}
\label{sec:theorem}
Detailed theoretical results on FWER control and detection accuracy are presented in the Appendix, while the proofs of the theorems are provided in the Supplementary Material.

\subsection{Family-wise error rate control}
Recall that $U_n = \boldsymbol{G}^\top (Y - \hat{\eta}_0)/\sqrt{n}$ is the vector of marginal scores and the test statistic is the maxima of marginal scores. Under the null hypothesis, $Y - \hat{\eta}_0$ is asymptotically distributed as $\mathcal{N}(0, P)$, where $P$ is the projection matrix. Let $Z \sim \mathcal{N}(0, P)$, under some regularity assumptions, the following theorem claims that the proposed global test controls the size and consequently, the dBiRS algorithm controls the FWER.

\begin{thm}
    \label{thm:FWER}
    Assume that the following conditions (a)-(b) hold:\\
    (a) For $i = 1, \dots, n$, the function $w_i(x_0, x_1) = 1/a_i(x_0)\nu(x_1)$ has finite gradient on a neighborhood of $(\phi, \eta_i)$;\\
    (b) $\log^4(p)/n \rightarrow 0$ as $n \rightarrow \infty$.\\
    Then under the null hypothesis, with probability one, we have 
    \begin{equation*}
        \abs{\mathbb{P}(\norm{U_n}_{\infty} > \tilde{c}(\alpha)) - \alpha} \rightarrow 0,
    \end{equation*}
    and consequently, the dBiRS procedure controls the FWER.
\end{thm}

Condition (a) is used for consistently estimating $\Sigma$ by $\hat{\Sigma}$ and is common in asymptotic analysis for GLM. Condition (b) indicates that the data dimension $p$ can grow exponentially in $n$. Recall that the sBiRS algorithm begins with the global test, then the sBiRS procedure controls the FWER while the size of the global test is controlled. Furthermore, since we perform a sBiRS procedure for blocks in dBiRS algorithm, the FWER of dBiRS is the same as sBiRS, which indicates that the dBiRS procedure also controls the FWER.

\subsection{Detection accuracy analysis}
Now we concentrate on the accuracy of the detection algorithm in terms of the proportion of true discoveries and false discoveries. Before the rigorous power analysis of our algorithm, we assume genotype data is bounded and centralized and introduce some notations. Given a truncation parameter $s$, we can split the global region into several continuous segments such that the length of each segment is less than $2^s$ and at least one segment has a length greater than $2^{s - 1}$. Note that the split is not unique, and let $\mathcal{S} = \left\{ I_1, \dots, I_S \right\}$ be the set containing these segments of any split. The regions $I_1, \dots, I_S$ are neighboring and non-overlapped satisfying $\cup_{j=1}^K I_j = \{1,\ldots,p\}$. For a true signal region $I_r^*, r=1,\ldots,\ell$, we refer to $I_{i_r}, \dots, I_{i_r+k_r}$ as the minimum cover of $I_r^*$ in $\mathcal{S}$ if $I^* \subset \cup_{j=0}^k I_{i+j}$ and  $\abs{\cup_{j = 0}^{k - 1}I_{i+j}} < \abs{I^*} \leq \abs{\cup_{j=0}^k I_{i+j}}$. Finally, without loss of generality, we suppose that $I_1^*, \dots, I_{\ell}^*$ with lengths $p_1, \dots, p_{\ell}$ have decayed signals, that is, $\norm{\mu_{I_1^*}}_{\infty} \geq \norm{\mu_{I_2^*}}_{\infty} \geq \cdots \geq \norm{\mu_{I_{\ell}^*}}_{\infty}$, where $\mu = \mathbb{E}(U)/\sqrt{n}$.

Providing a decay signal strength assumption for the true signal regions, we have the following theorem which gives the approximation results of true discoveries.
\begin{thm}
    \label{thm:TPR}
    Assume that the condition (a)-(b) in Theorem \ref{thm:FWER} hold, and further assume that\\ 
    (c) There exists a sufficiently large constant $C_1 > 0$, for $u = 0, \dots, k_r$, $r = 1, \dots, \ell$, 
    $$\norm{\mu_{I_{i_r+u}}}_{\infty} \geq C_1\left[\log\left\{ \left(p - \sum_{v=1}^{r-1}p_v\right)n \right\}/n\right]^{1/2}.$$\\
    Let $P_{\ell}$ denotes the probability that our dBiRS algorithm detects all $\ell$ signal regions, when $n \rightarrow \infty$, 
    $$P_{\ell} \geq 1 - 4\log(p+1)/n^2 \rightarrow 1.$$
\end{thm}

The result of this theorem states that in both the balanced and decaying signal settings, the dBiRS-detected signal segments can consistently cover the true signal regions. Condition (c) implies that the lower bound of $\norm{\mu_{I_1^*}}_2^2$ for consistent detection is of order $\log(p)\abs{I_1^*}/2^s/n$. The lower bound in Theorem 3 of Q-SCAN \citep{Lin20} is weaker than this, however, they assume that the maximum eigenvalue of $\hat{\Sigma}_{I_1^*}$ has a fixed upper bound, which is overly restricted. Furthermore, to the best of our knowledge, condition (c) is the first to allow signal decay settings in GLM model, while the existing work \citep{Cai10, Lin20} assumes that the signal strengths of all regions have a lower bound and \cite{zhang2023binary} adopts this decay setting under the two-sample test framework. 

Next, we focus on the false discovery rate of the dBiRS algorithm and illustrate the consistent detection property of our algorithm. To begin with, we introduce the Jaccard index to quantify the similarity between two regions. Specifically, we define the Jaccard index between sets $I_1$ and $I_2$ as 
$$J(I_1, I_2) = \abs{I_1 \cap I_2}/\abs{I_1 \cup I_2}.$$

Recall that $\mathcal{I}^* = \left\{ I_1^*, \dots, I_{\ell}^* \right\}$ are the set of true signal regions and $\hat{\mathcal{I}} = \left\{ \hat{I}_1, \dots, \hat{I}_{\ell'} \right\}$ are the estimated signal regions. For a signal region $I^*_i$, we define that it is consistently detected if for some $\eta(p) = o(1)$, there exists $\hat{I}_{j_i} \in \hat{\mathcal{I}}$ such that 
$$\prob\left\{ J(\hat{I}_{j_i}, I^*_i) \geq 1 - \eta(p) \right\} \rightarrow 1, $$ 
as $p \rightarrow \infty$. Note that, even if every signal region is consistently detected, there still could be some additional regions that are incorrectly detected.  Let $\tilde{I} = \hat{I}-I^{*} $ as the set of wrongly detected regions, and we want such regions to be ignorable relative to the true signal regions, 
i.e., $\abs{\tilde{I}}/\abs{I^*} \rightarrow 0$. 
Then we say that a detection procedure consistently detected all the true signal regions, if for a sequence of $\eta_j(p) = o(1), j = 1, \dots, \ell$ and some $\eta(p) = o(1)$, there exists $\hat{I}_{j_1}, \dots, \hat{I}_{j_\ell} \in \hat{\mathcal{I}}$ such that 
$$\prob\left[\left\{ \abs{\tilde{I}}/\abs{I^*} \leq \eta(p) \right\} \cap A\right] \rightarrow 1$$, 
where $A = \cap_{i=1}^{\ell}\left\{ J(\hat{I}_{j_i}, I_i^*) \geq 1 - \eta_j(p) \right\}$. 

Under some mild conditions concentrating on the structure of the covariance matrix $\Sigma$ and the distribution of true signal regions, we derive the detection consistency of dBiRS algorithm in the following theorem.

\begin{thm}
\label{thm:FDR}

Assume that conditions (a)-(c) hold. We further assume that\\
(d) There exists certain truncation parameter $s$ that results in the set of regions $\mathcal{S} = \left\{ I_1, \dots, I_S \right\}$ such that for some $\tilde{\alpha} > 2\alpha$, $\mathcal{O} = \left\{1, 2, \dots, S\right\}$, 
\begin{equation*}
    D(s, R) = \sup_{i_1, \dots, i_R \in \mathcal{O}}\frac{\mathbb{P}(\cap_{j=1}^R A_{i_j})}{\prod_{j=1}^R\mathbb{P}(A_{i_j})} \leq \frac{1}{(2\tilde{\alpha})^R},
\end{equation*}
where the events $A_i = \left\{ \norm{U_n(I_i)}_{\infty} \geq c_{I_i}(\alpha) \right\}$ and $c_{I_i}(\alpha)$ is the $1 - \alpha$ quantile of $\norm{U_n(I_i)}_{\infty}$;\\
(e) The true signal regions are well-separated in the sense that $Gap_{\min}  \geq L_{\max}$, where $L_{\max}$ is the maximum length of all true continuous signal regions and $G_{\min}$ is the minimum length of the gaps between any two true continuous signal regions;\\
(f) There exists a truncation parameter $s$ satisfies (d) and $s = o\left\{ \log_2(L_{\min}/\log\ell) \right\}$;\\
(g) There exists $r = o(L_{\min})$ such that for any index $i \in \cup_{l=1}^{\ell} I^*_l$, the non-signal point $j$ satisfies that if $\abs{j - i} > r$, then 
\begin{equation*}
    \abs{\boldsymbol{G}_{\cdot j}^\top \boldsymbol{G}\beta}/\sqrt{n} = o(1).
\end{equation*}

Denote $h_i = \lceil \abs{I_i^*}/2^s \rceil$ as the cardinality of the minimum cover (from $\mathcal{S}$) of $I^*_i$, $i = 1, \dots, \ell$ and $\tilde{r} = \lceil r/2^s \rceil$. Given a significance level $\alpha$, when $n, p \rightarrow \infty$, for any sequence of integers $R_i = o(h_i)$ and $R_i/\log \ell \rightarrow \infty$, $i = 1, \dots, \ell$, there exists $\hat{I}_{j_1}, \dots, \hat{I}_{j_\ell} \in \hat{\mathcal{I}}$ such that, 
\begin{equation*}
\prob\left[ \bigcap\limits_{i = 1}^{\ell} \left\{ J\left( I_i^{*},\hat{I}_{j_i} \right) \geq 1 - \eta_i \right\} \right] \geq 1 - \delta,
\end{equation*}
where $\eta_i = (R_i + 2\tilde{r})/(R_i+h_i+1) \rightarrow 0$ and $\delta = 4\log(p + 1)/n^2 + \sum_{i=1}^{\ell} \left(\alpha/2\tilde{\alpha}\right)^{R_i} \rightarrow 0$.

Let $A = \cap_{i=1}^{\ell}\left\{ J(\hat{I}_{j_i}, I_i^*) \geq 1 - \eta_j \right\}$, we have that for some integer $R_0 = O(K_1)$ and $R_0 = o(\sum_{i=1}^{\ell}h_i)$,
\begin{equation}
\label{eq:FinalP}
\prob\left[\left\{ \abs{\tilde{I}}/\abs{I^*} \leq \eta_0 \right\} \cap A\right] \geq \left(1 - \delta\right)\left[ 1 - C_2\left\{\frac{2R_0\alpha}{(R_0 - K_1)\tilde{\alpha}}\right\}^{R_0 - K_1} \right] \rightarrow 1,
\end{equation}
where $\eta_0 = R_0/\sum_{i=1}^\ell h_i \rightarrow 0$, $K_1 \leq K$ is the number of blocks which have signals and $C_2$ is a constant.
\end{thm}

To appreciate this result, we see that $1-4\log(p + 1)/n^2$ quantifies the probability that there exist estimated signal regions simultaneously covering the true ones as Theorem \ref{thm:TPR} asserts, while $(\alpha/2\tilde{\alpha})^{R_i}$ indicates how many non-signal segments in $\mathcal{S}$  may be falsely included neighboring to $I_i^*$ resulting from the re-search procedure. The term $1 - C_2\left\{\frac{2R_0\alpha}{(R_0 - K_1)\tilde{\alpha}}\right\}^{R_0 - K_1}$ in inequality \eqref{eq:FinalP} is the lower bound of the probability that $R_0$ additional non-signal segments are falsely detected and it relates to the number of blocks $K$. 

Theorem \ref{thm:FDR} demonstrates that the consistent detection property can be achieved by the dBiRS algorithm under mild conditions. Condition (d) relaxes the common M-dependence assumption in \cite{Lin20} and only requires a ``weak dependence'' assumption under which the Jaccard index consistency of the dBiRS algorithm is still guaranteed. It is straightforward to verify that the M-dependence satisfies condition (d) if we choose the truncation parameter $s \ge \log_2 M$. Moreover, condition (d) includes a larger class of covariance matrices. For instance, the covariance $\Sigma = \left\{ \left(1 + \abs{j - k}\right)^{-\rho} \right\}_{j,k=1}^p$, $\rho > 1/2$ and $\Sigma = \left\{\theta^{\abs{j - k}}\right\}_{j,k=1}^p, \theta < 1$ satisfy condition (d) but is not M-dependent. Condition (e) imposes the requirement on the distances among the signal regions. The constraint $2^s = o\left(L_{\min}/\log\ell\right)$ in condition (f) guarantees that the union set of the minimum cover of a true signal region is as short as possible. Conditions (e) and (f) are weaker than those in scan-based methods. Specifically, the scan-based method \cite{Lin20} assumes that $L_{\min}/\log p \rightarrow \infty$, while the dBiRS algorithm has less restriction on the minimum length of signal regions. For instance, consider $L_{\min} = \log p$ and $\ell = \log p$, then we can select $2^s = \sqrt{\log p}$ for consistent detection. This suggests that the dBiRS algorithm can detect shorter (and unbalanced) signal regions than the scan-based method that requires $L_{\min}/\log p\rightarrow \infty$ and hence can be more accurate. Moreover, the dBiRS algorithm does not have restrictions on the maximum length of signal regions that the scan method demands and can deal with signal regions with a length of (polynomial) order $p$. For illustration, let the two signal regions are $I^*_1 = \left\{1, \dots, \lfloor p/4 \rfloor \right\}$ and $I^*_2 = \left\{\lfloor 3/4p \rfloor, \dots, p\right\}$. According to Theorem \ref{thm:FDR}, dBiRS algorithm can consistently detect the signal regions, whereas the consistency of Q-SCAN \citep{Lin20} requires $\log L_{\max}/\log p\rightarrow 0$. Condition (g) requires the correlations between signal regions and non-signal points to decay with the distance, which is a standard assumption in this field.

\bigskip
\begin{center}
{\large\bf SUPPLEMENTARY MATERIAL}
\end{center}


\begin{description}
\item[dBiRS\_supp] Additional technical assumptions, lemmas and proofs of the theoretical results for ``Computationally Efficient Whole-Genome Signal Region Detection for Quantitative and Binary Traits''. (.pdf file)

\end{description}
\small
\bibliographystyle{agsm}

\bibliography{Bibliography-MM-MC}

@article{Xue20,
	Author = {Kaijie Xue and Fang Yao},
	Doi = {10.1214/19-AOS1848},
	Journal = {The Annals of Statistics},
	Number = {3},
	Pages = {1304 -- 1328},
	Publisher = {Institute of Mathematical Statistics},
	Title = {{Distribution and correlation-free two-sample test of high-dimensional means}},
	Volume = {48},
	Year = {2020}}

@article{Cai10,
	Author = {X. Jessie Jeng and T. Tony Cai and Hongzhe Li},
	Doi = {10.1198/jasa.2010.tm10083},
	Eprint = {https://doi.org/10.1198/jasa.2010.tm10083},
	Journal = {Journal of the American Statistical Association},
	Note = {PMID: 23543902},
	Number = {491},
	Pages = {1156-1166},
	Publisher = {Taylor & Francis},
	Title = {Optimal Sparse Segment Identification With Application in Copy Number Variation Analysis},
	Volume = {105},
	Year = {2010}
}

@article{Lin20,
	Author = {Zilin Li and Yaowu Liu and Xihong Lin},
	Doi = {10.1080/01621459.2020.1822849},
	Eprint = {https://doi.org/10.1080/01621459.2020.1822849},
	Journal = {Journal of the American Statistical Association},
	Number = {0},
	Pages = {1-12},
	Publisher = {Taylor & Francis},
	Title = {Simultaneous Detection of Signal Regions Using Quadratic Scan Statistics With Applications to Whole Genome Association Studies},
	Volume = {0},
	Year = {2020}}

@article{Visscher17,
	Author = {Visscher, Peter M. and Wray, Naomi R. and Zhang, Qian and others},
	Journal = {The American Journal of Human Genetics},
	Number = {1},
	Pages = {5-22},
	Title = {10 Years of GWAS Discovery: Biology, Function, and Translation},
	Volume = {101},
	Year = {2017}}

@article{Naus1982,
	Author = {Joseph I. Naus},
	Doi = {10.1080/01621459.1982.10477783},
	Eprint = {https://www.tandfonline.com/doi/pdf/10.1080/01621459.1982.10477783},
	Journal = {Journal of the American Statistical Association},
	Number = {377},
	Pages = {177-183},
	Publisher = {Taylor & Francis},
	Title = {Approximations for Distributions of Scan Statistics},
	Volume = {77},
	Year = {1982}}

@article{liu2010versatile,
  title={A versatile gene-based test for genome-wide association studies},
  author={Liu, Jimmy Z and Mcrae, Allan F and Nyholt, Dale R and others},
  journal={The American Journal of Human Genetics},
  volume={87},
  number={1},
  pages={139--145},
  year={2010},
  publisher={Elsevier}
}

@article{bakshi2016fast,
  title={Fast set-based association analysis using summary data from GWAS identifies novel gene loci for human complex traits},
  author={Bakshi, Andrew and Zhu, Zhihong and Vinkhuyzen, Anna AE and others},
  journal={Scientific reports},
  volume={6},
  number={1},
  pages={1--9},
  year={2016},
  publisher={Springer}
}

@article{he2021identification,
  title={Identification of putative causal loci in whole-genome sequencing data via knockoff statistics},
  author={He, Zihuai and Liu, Linxi and Wang, Chen and others},
  journal={Nature communications},
  volume={12},
  number={1},
  pages={3152},
  year={2021a},
  publisher={Nature Publishing Group UK London}
}

@article{he2021genome,
  title={Genome-wide analysis of common and rare variants via multiple knockoffs at biobank scale, with an application to Alzheimer disease genetics},
  author={He, Zihuai and Le Guen, Yann and Liu, Linxi and others},
  journal={The American Journal of Human Genetics},
  volume={108},
  number={12},
  pages={2336--2353},
  year={2021b},
  publisher={Elsevier}
}

@article{Olshen04,
    author = {Olshen, Adam B. and Venkatraman, E. S. and Lucito, Robert and Wigler, Michael},
    title = "{Circular binary segmentation for the analysis of array‐based DNA copy number data}",
    journal = {Biostatistics},
    volume = {5},
    number = {4},
    pages = {557-572},
    year = {2004},
    month = {10},
    issn = {1465-4644},
    doi = {10.1093/biostatistics/kxh008}
}

@article{Zhang10,
 ISSN = {00063444, 14643510},
 author = {Nancy R. Zhang and David O. Siegmund and Hanlee Ji and Jun Z. Li},
 journal = {Biometrika},
 number = {3},
 pages = {631--645},
 publisher = {[Oxford University Press, Biometrika Trust]},
 title = {Detecting simultaneous changepoints in multiple sequences},
 urldate = {2024-01-15},
 volume = {97},
 year = {2010}
}

@article{kircher2014general,
  title={A general framework for estimating the relative pathogenicity of human genetic variants},
  author={Kircher, Martin and Witten, Daniela M and Jain, Preti and O'roak, Brian J and Cooper, Gregory M and Shendure, Jay},
  journal={Nature genetics},
  volume={46},
  number={3},
  pages={310--315},
  year={2014},
  publisher={Nature Publishing Group US New York}
}

@article{Li2020dynamic,
    author = {Li, Xihao and Li, Zilin and Zhou, Hufeng and others},
    title = {Dynamic incorporation of multiple in silico functional annotations empowers rare variant association analysis of large whole-genome sequencing studies at scale},
    journal = {Nature genetics},
    year = {2020}, 
    volume = {52},  
    pages = {969--983}
}

@article{zhou2023favor,
  title={FAVOR: functional annotation of variants online resource and annotator for variation across the human genome},
  author={Zhou, Hufeng and Arapoglou, Theodore and Li, Xiao and others},
  journal={Nucleic Acids Research},
  volume={51},
  number={D1},
  pages={D1300--D1311},
  year={2023},
  publisher={Oxford University Press}
}

@article{Li2022nature,
    author = {Li, Zilin and Li, Xihao and Zhou, Hufeng and others},
    title = {A framework for detecting noncoding rare-variant associations of large-scale whole-genome sequencing studies},
    journal = {Nature Methods},
    year = {2022}, 
    volume = {19},  
    pages = {1599-1611}
}

@article{MORGENTHALER200728,
    title = {A strategy to discover genes that carry multi-allelic or mono-allelic risk for common diseases: A cohort allelic sums test (CAST)},
    journal = {Mutation Research/Fundamental and Molecular Mechanisms of Mutagenesis},
    volume = {615},
    number = {1},
    pages = {28-56},
    year = {2007},
    author = {Stephan Morgenthaler and William G. Thilly}
}

@article{Madsen2009,
    author = {Madsen, Bo Eskerod AND Browning, Sharon R.},
    journal = {PLOS Genetics},
    title = {A Groupwise Association Test for Rare Mutations Using a Weighted Sum Statistic},
    year = {2009},
    volume = {5},
    pages = {1-11},
    number = {2}
}

@article{Wu2011AJHG,
    author = {Wu, Micheal C. and Lee, Seunggeun and Cai, Tianxi and Li, Yun and Micheal Boehnke and Lin, Xihong},
    title = {Rare-Variant Association Testing for Sequencing Data With the Sequence Kernel Association Test},
    journal = {The American Journal of Human Genetics},
    year = {2011}, 
    volume = {89}, 
    number = {1},
    pages = {82-93}
}

@article{cosi2014,
  title={Cosi2: an efficient simulator of exact and approximate coalescent with selection},
  author={Ilya Shlyakhter Pardis C. Sabeti and Stephen F. Schaffner},
  journal={Bioinformatics},
  volume={30},
  number={23},
  pages={3427--3429},
  year={2014}
}

@article{zhang2023binary,
      title={Binary and Re-search Signal Region Detection in High Dimensions}, 
      author={Wei Zhang and Fan Wang and Fang Yao},
      year={2023},
      eprint={2305.08172},
      archivePrefix={arXiv},
      primaryClass={stat.ME},
    journal={arXiv preprint arXiv:2305.08172}
}

@article{cai2022distributed,
author = {T. Tony Cai and Hongji Wei},
title = {{Distributed adaptive Gaussian mean estimation with unknown variance: Interactive protocol helps adaptation}},
volume = {50},
journal = {The Annals of Statistics},
number = {4},
publisher = {Institute of Mathematical Statistics},
pages = {1992 -- 2020},
year = {2022},
doi = {10.1214/21-AOS2167},
URL = {https://doi.org/10.1214/21-AOS2167}
}

@article{li2022simultaneous,
  title={Simultaneous detection of signal regions using quadratic scan statistics with applications to whole genome association studies},
  author={Li, Zilin and Liu, Yaowu and Lin, Xihong},
  journal={Journal of the American Statistical Association},
  volume={117},
  number={538},
  pages={823--834},
  year={2022},
  publisher={Taylor \& Francis}
}

@article{kanki2020creb,
  title={CREB coactivator CRTC2 plays a crucial role in endothelial function},
  author={Kanki, Hideaki and Sasaki, Tsutomu and Matsumura, Shigenobu and Kawano, Tomohiro and Todo, Kenichi and Okazaki, Shuhei and Nishiyama, Kumiko and Takemori, Hiroshi and Mochizuki, Hideki},
  journal={Journal of Neuroscience},
  volume={40},
  number={49},
  pages={9533--9546},
  year={2020},
  publisher={Soc Neuroscience}
}

@article{savage2018genome,
  title={Genome-wide association meta-analysis in 269,867 individuals identifies new genetic and functional links to intelligence},
  author={Savage, Jeanne E and Jansen, Philip R and Stringer, Sven and Watanabe, Kyoko and Bryois, Julien and De Leeuw, Christiaan A and Nagel, Mats and Awasthi, Swapnil and Barr, Peter B and Coleman, Jonathan RI and others},
  journal={Nature genetics},
  volume={50},
  number={7},
  pages={912--919},
  year={2018},
  publisher={Nature Publishing Group US New York}
}

@article{mandal2018brwd1,
  title={BRWD1 orchestrates epigenetic landscape of late B lymphopoiesis},
  author={Mandal, Malay and Maienschein-Cline, Mark and Maffucci, Patrick and Veselits, Margaret and Kennedy, Domenick E and McLean, Kaitlin C and Okoreeh, Michael K and Karki, Sophiya and Cunningham-Rundles, Charlotte and Clark, Marcus R},
  journal={Nature communications},
  volume={9},
  number={1},
  pages={3888},
  year={2018},
  publisher={Nature Publishing Group UK London}
}

@article{davies2018study,
  title={Study of 300,486 individuals identifies 148 independent genetic loci influencing general cognitive function},
  author={Davies, Gail and Lam, Max and Harris, Sarah E and Trampush, Joey W and Luciano, Michelle and Hill, W David and Hagenaars, Saskia P and Ritchie, Stuart J and Marioni, Riccardo E and Fawns-Ritchie, Chloe and others},
  journal={Nature communications},
  volume={9},
  number={1},
  pages={2098},
  year={2018},
  publisher={Nature Publishing Group UK London}
}

@article{herold2016family,
  title={Family-based association analyses of imputed genotypes reveal genome-wide significant association of Alzheimer’s disease with OSBPL6, PTPRG, and PDCL3},
  author={Herold, Christine and Hooli, Basavaraj V and Mullin, Kristina and Liu, Tian and Roehr, Johannes T and Mattheisen, Manuel and Parrado, Antonio R and Bertram, Lars and Lange, Christoph and Tanzi, Rudolph E},
  journal={Molecular psychiatry},
  volume={21},
  number={11},
  pages={1608--1612},
  year={2016},
  publisher={Nature Publishing Group}
}

@article{ripke2011schizophrenia,
  title={Schizophrenia Psychiatric Genome-Wide Association Study (GWAS) Consortium Genome-wide association study identifies five new schizophrenia loci},
  author={Ripke, S and Sanders, AR and Kendler, KS and Levinson, DF and Sklar, P and Holmans, P and Lin, DY and Duan, J and Ophoff, RA and Andreassen, OA and others},
  journal={Nat. Genet},
  volume={43},
  pages={969--976},
  year={2011}
}

@article{zhang2023brain,
  title={A brain-wide genome-wide association study of candidate quantitative trait loci associated with structural and functional phenotypes of pain sensitivity},
  author={Zhang, Li and Pan, Yiwen and Huang, Gan and Liang, Zhen and Li, Linling and Zhang, Min and Zhang, Zhiguo},
  journal={Cerebral Cortex},
  volume={33},
  number={11},
  pages={7297--7309},
  year={2023},
  publisher={Oxford University Press}
}

@article{king2013topoisomerases,
  title={Topoisomerases facilitate transcription of long genes linked to autism},
  author={King, Ian F and Yandava, Chandri N and Mabb, Angela M and Hsiao, Jack S and Huang, Hsien-Sung and Pearson, Brandon L and Calabrese, J Mauro and Starmer, Joshua and Parker, Joel S and Magnuson, Terry and others},
  journal={Nature},
  volume={501},
  number={7465},
  pages={58--62},
  year={2013},
  publisher={Nature Publishing Group UK London}
}

@article{madabhushi2015activity,
  title={Activity-induced DNA breaks govern the expression of neuronal early-response genes},
  author={Madabhushi, Ram and Gao, Fan and Pfenning, Andreas R and Pan, Ling and Yamakawa, Satoko and Seo, Jinsoo and Rueda, Richard and Phan, Trongha X and Yamakawa, Hidekuni and Pao, Ping-Chieh and others},
  journal={Cell},
  volume={161},
  number={7},
  pages={1592--1605},
  year={2015},
  publisher={Elsevier}
}

@article{tiwari2012target,
  title={Target genes of Topoisomerase II$\beta$ regulate neuronal survival and are defined by their chromatin state},
  author={Tiwari, Vijay K and Burger, Lukas and Nikoletopoulou, Vassiliki and Deogracias, Ruben and Thakurela, Sudhir and Wirbelauer, Christiane and Kaut, Johannes and Terranova, Remi and Hoerner, Leslie and Mielke, Christian and others},
  journal={Proceedings of the National Academy of Sciences},
  volume={109},
  number={16},
  pages={E934--E943},
  year={2012},
  publisher={National Acad Sciences}
}

@article{wang2020genome,
  title={Genome-wide interaction analysis of pathological hallmarks in Alzheimer's disease},
  author={Wang, Hui and Yang, Jingyun and Schneider, Julie A and De Jager, Philip L and Bennett, David A and Zhang, Hong-Yu},
  journal={Neurobiology of aging},
  volume={93},
  pages={61--68},
  year={2020},
  publisher={Elsevier}
}

@article{lee2018gene,
  title={Gene discovery and polygenic prediction from a genome-wide association study of educational attainment in 1.1 million individuals},
  author={Lee, James J and Wedow, Robbee and Okbay, Aysu and Kong, Edward and Maghzian, Omeed and Zacher, Meghan and Nguyen-Viet, Tuan Anh and Bowers, Peter and Sidorenko, Julia and Karlsson Linn{\'e}r, Richard and others},
  journal={Nature genetics},
  volume={50},
  number={8},
  pages={1112--1121},
  year={2018},
  publisher={Nature Publishing Group US New York}
}

@article{okbay2022polygenic,
  title={Polygenic prediction of educational attainment within and between families from genome-wide association analyses in 3 million individuals},
  author={Okbay, Aysu and Wu, Yeda and Wang, Nancy and Jayashankar, Hariharan and Bennett, Michael and Nehzati, Seyed Moeen and Sidorenko, Julia and Kweon, Hyeokmoon and Goldman, Grant and Gjorgjieva, Tamara and others},
  journal={Nature genetics},
  volume={54},
  number={4},
  pages={437--449},
  year={2022},
  publisher={Nature Publishing Group}
}

@article{meier2019genetic,
  title={Genetic variants associated with anxiety and stress-related disorders: a genome-wide association study and mouse-model study},
  author={Meier, Sandra M and Trontti, Kalevi and Purves, Kirstin L and Als, Thomas Damm and Grove, Jakob and Laine, Mikaela and Pedersen, Marianne Gi{\o}rtz and Bybjerg-Grauholm, Jonas and B{\ae}kved-Hansen, Marie and Sokolowska, Ewa and others},
  journal={JAMA psychiatry},
  volume={76},
  number={9},
  pages={924--932},
  year={2019},
  publisher={American Medical Association}
}

@article{pasman2022genetic,
  title={Genetic risk for smoking: disentangling interplay between genes and socioeconomic status},
  author={Pasman, Jo{\"e}lle A and Demange, Perline A and Guloksuz, Sinan and Willemsen, AHM and Abdellaoui, Abdel and Ten Have, Margreet and Hottenga, Jouke-Jan and Boomsma, Dorret I and de Geus, Eco and Bartels, Meike and others},
  journal={Behavior genetics},
  volume={52},
  number={2},
  pages={92--107},
  year={2022},
  publisher={Springer}
}

@article{hollins2016microrna,
  title={MicroRNA: Small RNA mediators of the brains genomic response to environmental stress},
  author={Hollins, Sharon L and Cairns, Murray J},
  journal={Progress in neurobiology},
  volume={143},
  pages={61--81},
  year={2016},
  publisher={Elsevier}
}

@article{manolio2009finding,
  title={Finding the missing heritability of complex diseases},
  author={Manolio, Teri A and Collins, Francis S and Cox, Nancy J and Goldstein, David B and Hindorff, Lucia A and Hunter, David J and McCarthy, Mark I and Ramos, Erin M and Cardon, Lon R and Chakravarti, Aravinda and others},
  journal={Nature},
  volume={461},
  number={7265},
  pages={747--753},
  year={2009},
  publisher={Nature Publishing Group UK London}
}
\end{document}